\documentstyle[12pt]{article}
\newcommand{\be}{\begin{equation}}
\newcommand{\ee}{\end{equation}}
\begin{document}
\begin{titlepage}
\title{
{\bf Geometrical string and dual spin systems
\footnote{Dedicated to Professor
Herbert Wagner at the occasion of his 60th birthday}}
}%title ends
{\bf
\author{
 G.K. Savvidy\\
 Physics Department, University of Crete \\
 71409 Iraklion, Crete, Greece ; \\
 Yerevan Physics Institute, 375036 Yerevan, Armenia\\
  \vspace{1cm}\\
  K.G. Savvidy\\
 Princeton University, Department of Physics\\
 P.O.Box 708, Princeton, New Jersey 08544, USA \\
 \vspace{1cm}\\
 F.J. Wegner\\
 Institut f\"ur Theoretische Physik\\
 Ruprecht-Karls-Universit\"at Heidelberg\\
 Philosophenweg 19, D-69120 Heidelberg, Germany
}%author ends
}
\date{}%in order NOT to write the date
\maketitle
\begin{abstract}
\noindent

We are able to perform the duality transformation of the spin
system which was found before as a lattice realization of the
string with linear action. In four and higher dimensions this
spin system can be described in terms of a two-plaquette gauge
Hamiltonian. The duality transformation is constructed in
geometrical and algebraic language. The dual Hamiltonian
represents a new type of spin system with local gauge
invariance. At each vertex $\xi$ there are $d(d-1)/2$
Ising spins $\Lambda_{\mu,\nu}= \Lambda_{\nu,\mu}$,
$\mu \neq \nu = 1,..,d$ and one Ising spin $\Gamma$
on every link $(\xi,\xi +e_{\mu})$. For the frozen spin
$\Gamma \equiv 1$
the dual Hamiltonian factorizes into $d(d-1)/2$ two-dimensional
Ising ferromagnets and into antiferromagnets in the case
$\Gamma \equiv -1$. For fluctuating $\Gamma$ it is a sort of
spin glass system with local gauge invariance.
The generalization to $p$-branes is given.

\end{abstract}
\thispagestyle{empty}
\end{titlepage}
\pagestyle{empty}

\section{Introduction}
\subsection{String with linear action}
The strong coupling expansion of the lattice gauge theories and
topological classification of the QCD diagrams, together with
the last progress in solving two-dimensional QCD, put forward
the ultimate conjunction that QCD is exactly equivalent to a
string theory \cite{gross}. One can also expect a new progress in
understanding of the QCD string in four dimensions
\cite{douglas,matytsin}.
In this approach one should derive the
relevant string theory directly from properly regularized QCD
path integrals.

In the complementary approach \cite{durhuus,david,gross1,ambjorn}
one can try to derive the relevant
string Lagrangian from alternative principles such as
\cite{savvidy2} :
\vspace{.3cm}

$\alpha )$ natural coincidence of the string transition amplitude
with the usual Feynman path amplitude for long space-time strips
and

$\beta )$ the continuity principle for the string amplitudes.
\vspace{.3cm}

The partition function for the regularized string,
which can be derived from these principles has the form
\be
 Z(\beta) = \sum_{\{ M \}} \exp\{- \beta A(M) \} \label{Z},
\ee
where the summation is extended over all triangulated
surfaces $\{ M \}$ with the linear action $A(M)$
\be
A(M) = \sum_{<i,j>}  \lambda_{i,j}
\cdot \vert \pi - \alpha_{i,j} \vert \label{linear}
\ee
and $\alpha_{i,j}$~ is the angle between two neighbor
faces of $M$ having a common edge $<i,j>$ of the length
$\lambda_{i,j}$. The regularized string (\ref{Z}),
(\ref{linear})
is well defined in any dimensions and for arbitrary
topology of the surface $M$.

\subsection{Surfaces on the lattice}
In a recent paper \cite{savvidywegner} two of the authors
have introduced a spin
statistical system on the lattice, the $low$ temperature
expansion of which generates random surfaces with linear
action $A(M)$ (\ref{linear}). In this lattice implementation
of the linear string the corresponding Hamiltonian depends
on the way how we ascribe the weights on the self-intersection
edges.

Indeed as it is well known, the surfaces
of interface naturally appearing in the low temperature expansion
of the partition function have self-intersections, and it is
an important problem to find out a way to treat them properly.
In general one can define
corresponding weights at the self-intersection edges
in different ways
without spoiling the linear character of the theory
(\ref{linear}).

According to $\alpha$) and $\beta$)
one should count the
angles between plaquettes approaching the self-intersection
edges and then multiply the result by
$\lambda_{i,j}~~ $\cite{savvidy2}
\be
\lambda_{i,j}\cdot (\vert \pi - \alpha_{i,j} \vert +
\vert \pi - \beta_{i,j} \vert +....).
\ee
In \cite{savvidywegner} the authors  suggested to ascribe
the weigths equal to the length $\lambda_{i,j}$
multiplied by the $total$ number of the right
angles plaquettes approaching the edge $<i,j>$ .
One can also consider these weights with arbitrary
self-intersection coupling constant $k$
and observe that in the limit,
when the self-intersection coupling constant $k$ tends to
infinity, we end up with self-avoiding surfaces
\cite{savvidy3}.
It also became clear, that in the case of a three-dimensional
lattice the system drastically simplifies in the
opposite limit,  when the
self-intersection coupling constant $k$  tends to zero.
This simplification
is strongly connected with an additional symmetry
which appears in the system in this limit and
allows therefore to construct the dual Hamiltonian.

In the following we will first reconsider the
three-dimensional model and its dual, in section 2
surfaces with linear action in four dimensions are considered,
and in section 3 a formulation for general dimensions
will be given.

\subsection{Duality in three dimensions}

In the limit, when self-intersection coupling constant $k$
tends to zero, the additional symmetry and the positivity of the local
weights  allows to construct the  dual Hamiltonian in three dimensions
which generates random surfaces with linear action $A(M)$ now
in its $high$ temperature expansion.

In terms of Ising spin variables $\sigma_{\vec r}$
the Hamiltonian with self-intersection
coupling constant $k$ equal to zero has the form

\be
H^{3d}_{gonihedric} = - \sum_{\vec r,\vec \alpha,\vec \beta}
\sigma_{\vec r} \sigma_{\vec r+\vec \alpha}
\sigma_{\vec r+\vec \alpha+\vec \beta}
\sigma_{\vec r+\vec \beta} \label{3Dham}
\ee
where $\vec r$ is a three-dimensional vector on the lattice and $\vec \alpha$
,$\vec \beta$
are unit vectors parallel to the axes. We should stress that the Ising
spins in (\ref{3Dham}) are on the vertices of a
simple cubic lattice and are not on the links. The details of the
construction of the Hamiltonian (\ref{3Dham}) from the first
principles can be found in \cite{savvidywegner,savvidy3,savvidy4}.
In addition to the usual
symmetry group $Z_{2}$ the system (\ref{3Dham}) has a new
symmetry: one can independently flip
the spins on any combination of planes (spin layers)
of the three-dimensional lattice.

The dual Hamiltonian reads \cite{savvidy4}

\be
H^{3d}_{dual} = -\sum_{\xi} \Lambda^{\chi}(\xi)
\Lambda^{\chi}(\xi + \chi)
+ \Lambda^{\eta}(\xi) \Lambda^{\eta}(\xi + \eta)+
\Lambda^{\varsigma}(\xi) \Lambda^{\varsigma}(\xi + \varsigma) ,
\label{3Ddual}
\ee
where $\chi$,~$\eta$,~and $\varsigma$ are unit vectors
in the corresponding directions of the dual lattice and $\Lambda$'s are
one-dimensional irreducible representations of the forth order
Abelian group $G_{\xi} = \{ e, g_{\chi},g_{\eta},g_{\varsigma} \} $
\be
\Lambda^{\chi} = \{ 1,1,-1,-1 \},~~~ \Lambda^{\eta} =
\{ 1,-1,1,-1\},~~~\Lambda^{\varsigma}
 =\{1,-1,-1,1\}. \label{repr}
\ee
Spin $G_{\xi}$ should be ascribed to every vertex $\xi$ of the dual
lattice.
$\Lambda^{\chi}$ and $\Lambda^{\eta}$ may also be considered as
Ising spins, then
$\Lambda^{\varsigma} = \Lambda^{\chi} \Lambda^{\eta}$
and (\ref{repr}) describes
a modified Ashkin-Teller model \cite{ashkinteller}.
The partition function of the dual systems (\ref{3Ddual})
can be written in the form
\be
Z(\beta^{\star})
= \sum_{\{ G_{\xi} \}}
\exp\{-\beta^{\star}  H^{3d}_{dual} \}. \label{Zdual}
\ee
Both Hamiltonian (\ref{3Dham}) and (\ref{3Ddual})
generate random surfaces with
linear-gonihedric action $A(M)$ in $low$ and $high$
temperature expansion correspondingly.

One can compare this system with the 3d Ising ferromagnet
\be
H^{3d}_{Ising} = -\sum_{link} \sigma \sigma
\ee
and with it's dual Hamiltonian \cite{wegnermathphys}
\be
H^{3d}_{dual} = -\sum_{plaquetts}
\sigma \sigma\sigma \sigma
\ee
which generates random surfaces with area action.

\subsection{String interpretation}
There is an intensive study of the properties of
cluster boundaries of the $3d$ Ising ferromagnet
in order to
get the meaning of random surfaces on the lattice
governed by the area law \cite{marinari,caselle,cardy}.
It is also important to understand the phase structure of the
linear systems (\ref{3Dham}) and (\ref{3Ddual}) and to study the
properties of the ensemble of random surfaces generated
by this system. For that one can use the curvature representation
of the linear action $A(M)$ \cite{schneider} and then find
an equivalent representation of the partition
function (\ref{Z}) in terms of propagation
of the polygon-loops or string $P$ in a given direction.
The transition amplitude  is equal to \cite{savvidy4}
\be
\exp\{-k(P)-l(P) \}
\ee
where $k(P)$ is the total curvature of $P$ and $l(P)$ is
the length. The interaction term is proportional
to the overlapping length of the strings
\be
A_{int}=l(P_{1}\cap P_{2}).
\ee
In the first approximation ignoring  the interaction term
one can solve the system and see that it describes the
propagation of almost free 2d Ising fermions \cite{savvidy4}.
This is a strong indication that the system has second
order phase transition similar to 2d Ising ferromagnet.

Again comparing our system with $3d$ Ising
ferromagnet one can see that
the propagation of the  string came up with the
amplitude
\be
\exp\{-l(P) - s(P)\}
\ee
where $l(P)$ is the length and $s(P)$ is the area. The
interaction, which is now proportional to the overlapping
area
\be
A_{int} = s(P_{1} \cap P_{2})
\ee
is much stronger and probably cannot be
represented as a propagation of the free string
\cite{marinari,caselle}. There is an indication
\cite{rusakov} that $SU(N)$ lattice
gauge theories with one plaquette action
have similar representation,
the interaction is also
proportional to the overlapping area.

\subsection{Duality in higher dimensions}
Our aim is to construct
dual representations of the linear system (\ref{Z}),
(\ref{linear}) in four and higher dimensions.
There is a new complication compared with the
three-dimensional case where we have set the self-intersection
coupling constant $k$ equal to zero. Indeed if
in four dimensions we set the self-intersection coupling
constant $k$ equal to zero then the  Hamiltonian has one and
three plaquette terms \cite{savvidy3} and if we take $k=1$
\cite{savvidywegner} then it has one and two plaquette terms.
It is almost impossible to work with these complicated Hamiltonians.

It is crucial at this point to use the lattice version of the continuity
principle $\beta$) in the following form :
the self-intersection weight should be defined so that
after pairwise connection of plaquettes at the intersection
edge the weight remain the same.
Indeed on a three-dimensional lattice the self-intersection
of four plaquettes $uniquely$ decomposes into two pairs of
flat plaquettes in accordance with $\beta$).

To satisfy the continuity principle in four-
and higher dimensional lattices
we will take weights in the intersections, so that
plaquettes continuing straight across an edge
do not contribute to the energy and only
the number of plaquettes which do not continue
in a flat two-dimensional plane do count. In this case we
again have an additional symmetry and the positivity of the
local weights which allows to construct the dual Hamiltonian.
We will consider such a system in
the next section.

\section{Surfaces with linear action in four dimensions}

On a four-dimensional simple hypercubic lattice
one can have intersections of
two, four and six plaquettes at an edge \cite{savvidywegner}.
In accordance with $\beta$) the intersection of six
plaquettes contributes zero energy and can
be uniquely decomposed
into three flat pairs of plaquettes.
The intersection of four plaquettes
yields zero energy in the cases where the plaquettes lie in two
flat planes and with the energy equal to $\pi /2$ (see (\ref{linear}))
if a pair of plaquettes is left which does not lie in a plane.
Therefore a four-plaquette
intersection also uniquely decomposes into two flat planes in
the first case and into one flat plane and one "corner" in the
second case.

For this choice of the self-intersection weights the
Hamiltonian has the form
\be
H^{4d}_{gonihedric}=-\sum_{pairs~of~parallel~plaquettes}
(\sigma \sigma \sigma \sigma )_{P}^{~~||}
(\sigma \sigma \sigma \sigma )_{P} \label{4Dham}
\ee
where the summation is extended over all pairs of parallel
plaquettes in $3d$ cubes of the $4d$ lattice.
Here the Ising spins $\sigma_{\vec r,\vec r+\vec \alpha}$
are located on the center of the links
$(\vec r, \vec r + \vec \alpha)$ of the four dimensional lattice.
The method of construction of the Hamiltonian (\ref{4Dham})
from the given distribution of the surface weights is
similar to the one which was suggested in \cite{savvidywegner}.
{\it A posteriori} one can check that the low temperature expansion
of the partition function of this system
\be
Z(\beta)=\sum_{\{\sigma\}}\exp\{-\beta H^{4d}_{gonihedric}\}
\label{4Z}
\ee
is obtained by summation over all closed
surfaces $\{ M \}$ with the weight $\exp(-\beta A(M))$ where
the linear action $A(M)$ is given by the number of nonflat pairs of
plaquettes of the closed surface $M$.

To construct the dual Hamiltonian let us consider the high
temperature expansion of the partition function (\ref{4Z})
\be
Z(\beta) =\sum_{\{\sigma\}}\prod_{pairs~of~plaquettes}
\{\cosh\beta + \sinh\beta \cdot (\sigma \sigma \sigma \sigma )_{P}^{~~||}
(\sigma \sigma \sigma \sigma )_{P} \} \label{4Dparti}
\label{expan}
\ee
The first nonzero terms of the expansion are (i) a
$3d$ cube covered by the three pairs of plaquettes
and (ii) a loop of four pairs of plaquettes forming a torus
which can "live" only inside a $4d$ hypercube.
(These terms are those given by eqs. (\ref{constr2})
and (\ref{constr3})).
To have a natural geometrical
representation of all expansion terms,
one can associate the link $l$ connecting the centers of the
pairs of parallel plaquettes in a $3d$ cube. In this language
the first term is represented by the "hedgehog" or cross of three
perpendicular links inside the $3d$ cube and
we will use $\Gamma$ for this basic element. The
second term is represented by a loop in a $4d$ hypercube
and we will use $\Lambda$ for this element.

In order to have a complete description of high temperature configurations
$\{ \Sigma \}$ let us consider a $4d$ hypercube.
There are eight different configurations $\Gamma$
and six different configurations $\Lambda$ inside a $4d$
hypercube. It is convenient
to introduce therefore $\Gamma_{+\mu}$ and $\Gamma_{-\mu}$ ,
$\mu = 1,2,3,4$ , for these eight configurations $\Gamma$ and
$\Lambda_{\mu,\nu}= \Lambda_{\nu,\mu}$~$\mu \neq \nu $ for the six
configurations $\Lambda$. The indices naturally reflect
the positions of the elementary configurations inside the
$4d$ hypercube. Using these basic elements one can construct any
configuration $\Sigma$ of the high temperature expansion
(\ref{expan}). The energy of the given configuration $\Sigma$ is
equal to the total length of the links. The
configuration $\Gamma$ has the length equal to three
($(\tanh\beta)^{3}$ of  the high $T$ expansion) and a
$\Lambda$ has the length
equal to four $((\tanh\beta)^{4})$.

Let us ascribe now spin variables $\Gamma_{\pm \mu}$ and
$\Lambda_{\mu , \nu}$ to every vertex $\xi$ of the
dual lattice. They will assume the value $-1$
at the vertex $\xi$ in the cases when the $4d$ hypercube
is occupied by certain basic element $\Gamma_{\pm \mu}$
or  $\Lambda_{\mu\nu}$ and $+1$ if it is not.
To every center of the dual lattice we ascribe therefore
$14$ independent spins $\Gamma_{\pm \mu}$ and
$\Lambda_{\mu,\nu}(\xi)$.

There is an ambiguity in the definition of the $\Gamma_{\pm \mu}$.
One can ascribe configuration $\Gamma$ to any of two $4d$
hypercube to which the $3d$ cube with $\Gamma$ configuration
is the border. This corresponds to a local symmetry which
we will clarify later on (see \ref{sym4Da}).

Now one can construct the dual Hamiltonian
\be
H^{4d}_{dual} = \sum_{\xi}\sum^{4}_{\mu=1}H(\xi ,\xi + e_{\mu})
= -\sum_{\xi}\sum_{\nu \neq \mu}
\Lambda_{\mu,\nu}(\xi) \Lambda_{\mu,\nu}(\xi + e_{\mu})
\Gamma_{\mu}(\xi) \Gamma_{-\mu}(\xi + e_{\mu}), \label{4Ddual}
\ee
where $e_{\mu}$ is the unite vector in the corresponding
direction.

The Hamiltonian (\ref{4Ddual}) has a number of local gauge
symmetries: one can independently transform spins $\Gamma$
in neighbor vertices
\be
\Gamma_{\mu}(\xi) \rightarrow
-\Gamma_{\mu}(\xi), ~~~~~~~~\Gamma_{-\mu}(\xi+e_{\mu})
\rightarrow -\Gamma_{-\mu}(\xi+e_{\mu}),
\label{sym4Da}
\ee
and flip all spins at a given vertex
\be
(\Lambda(\xi),\Gamma(\xi)) \rightarrow -(\Lambda(\xi),\Gamma(\xi)).
\label{sym4Db}
\ee
Using local gauge invariance (\ref{sym4Da}) one can fix the
gauge
\be
\Gamma_{-\mu}(\xi) = 1 \label{gauge}
\ee
and therefore locate (or fix) the remaining spin
$\Gamma_{\mu}$ on the $link$ between two vertices
$\xi$ and $\xi +e_{\mu}$. It is convenient to use
notation $\Gamma(\xi, \xi +e_{\mu})$ for this
spin. The fixed gauge Hamiltonian has the form
\be
H^{4d}_{dual}= -\sum_{\xi}\sum_{\nu \neq \mu}
\Lambda_{\nu,\mu}(\xi)\cdot \Gamma(\xi,\xi + e_{\mu}) \cdot
\Lambda_{\mu,\nu}(\xi + e_{\mu}) \label{4Ddualfix}
\ee
The dual Hamiltonian (\ref{4Ddualfix}) is therefore a spin
system of six
Ising spins $\Lambda_{\mu,\nu}(\xi) = \Lambda_{\nu,\mu}(\xi)$,
$\mu \neq \nu = 1,2,3,4$ which are located on every $vertex$ $\xi$ of
the lattice and of one Ising spin  $\Gamma(\xi,\xi +e_{\mu})$
which is located on the center of every $link$ $(\xi,\xi +e_{\mu})$.

High temperature expansion of the
partition function

$$Z(\beta^{\star})=\sum_{\{\Lambda,\Gamma \}}\exp\{-\beta^{\star}
H^{4d}_{dual} \} $$
\be
= \sum_{ \{ \Lambda , \Gamma \}}
\prod_{\xi} \exp\{~ \beta^{\star} \sum_{i=2}^{4}
\Lambda_{1,i}(\xi) \cdot
\Gamma(\xi,\xi+e_{1})\cdot \Lambda_{1,i}(\xi +e_{1}) + ... \}
\label{Z4Ddual}
\ee
generate random surfaces with linear action $A(M)$.
Indeed, expanding (\ref{Z4Ddual}) for small $\beta^{\star}$
\be
Z(\beta^{\star})=\sum_{ \{ \Lambda , \Gamma \}}
\prod_{\xi}\prod_{i=2}^{4}
(\cosh\beta^{\star})^{4}
\{ 1+\tanh\beta^{\star} \cdot \Lambda_{1,i}(\xi) \cdot
\Gamma(\xi,\xi+e_{1})\cdot \Lambda_{1,i}(\xi +e_{1}) \}...
\ee
and then summing over $\Gamma(\xi,\xi+e_{1})$  and computing
the product over $i$ yields
$$ Z(\beta^{\star})=\sum_{\{ \Lambda \} }
\prod_{\xi}
(\cosh\beta^{\star})^{4}~ 2\cdot \{~ 1+(\tanh\beta^{\star})^{2}
\cdot~~~~~~~~~~~~~~~~~~~~~~~~~~~~$$ $$[ \Lambda_{1,2}(\xi)
\Lambda_{1,3}(\xi) \Lambda_{1,2}(\xi +e_{1})
\Lambda_{1,3}(\xi +e_{1})+ $$
$$\Lambda_{1,3}(\xi) \Lambda_{1,4}(\xi)
\Lambda_{1,3}(\xi +e_{1}) \Lambda_{1,4}(\xi +e_{1})+$$
\be
\Lambda_{1,4}(\xi) \Lambda_{1,2}(\xi)
\Lambda_{1,4}(\xi +e_{1}) \Lambda_{1,2}(\xi +e_{1})]~ \} ...
\ee
Opening the brackets and summing over spins $\Lambda$ one can
see that nonzero terms are those which correspond to
closed surfaces $M$ with linear action $A(M)$.

Both Hamiltonian (\ref{3Ddual}) and (\ref{4Ddual}) look
differently, but it is possible to rederive (\ref{3Ddual})
in the form which is similar to (\ref{4Ddual}),
(\ref{4Ddualfix}). For that let us
introduce three different Ising spins $\{\Lambda_{1},
\Lambda_{2},\Lambda_{3} \}$ in every vertex $\xi$, then
\be
H^{3d}_{dual} = \sum_{\xi}\sum^{3}_{i=1}H(\xi,\xi +e_{i})
= \sum_{i\neq j\neq k} \Lambda_{j}(\xi) \Lambda_{k}(\xi)
\Lambda_{j}(\xi+e_{i}) \Lambda_{k}(\xi+e_{i}) \label{L3Ddual}
\ee
and we observe the local gauge symmetry of the Hamiltonian
(\ref{L3Ddual})
\be
\Lambda_{1,2,3}(\xi) \rightarrow - \Lambda_{1,2,3}(\xi)\label{symm}
\ee

\section{Surfaces with linear action in high dimensions}
\subsection{The Hamiltonian}

In  \cite{savvidywegner} the authors have introduced also a
model of $(d-n)$-dimensional hypersurfaces on
a $d$-dimensional hypercubic lattice.
Such a hypersurface $M_{d-n}$ consists of a collection of
elementary hyperplaquettes $\Omega_{\alpha_{1}...\alpha_{n}}(\vec r)$ (all
$\alpha_{i}$ are different) with $\vec r$ on a simple
cubic lattice $Z^d$.
These are defined by $x_{\alpha_{i}} = r_{\alpha_{i}}$
and $r_{\alpha} \leq x_{\alpha} \leq r_{\alpha}+1$ for all
other $\alpha$.
As in \cite{savvidywegner} we introduce a variable
$U_{\alpha_{1}...\alpha_{n}}(\vec r)$, which assumes the value
$-1$ if the hyperplaquette belongs to the hypersurface $M_{d-n}$
and $+1$ if it does not.

The hypersurface $M_{d-n}$ shall be closed.
That is each $(d-n-1)$-dimensional hyperedge
$\Omega_{\alpha_{1}...\alpha_{n+1}}(\vec r)$ belongs totally to an
even number of $(d-n)$-dimensional hyperplaquettes of the
hypersurface $M_{d-n}$. This imposes the constraint
\be
\prod_{k=1}^{n+1}
 U_{\alpha_1...\alpha_{k-1}\alpha_{k+1}...\alpha_{n+1}}(\vec r)
U_{\alpha_1...\alpha_{k-1}\alpha_{k+1}...\alpha_{n+1}}
(\vec r-\vec e_{\alpha_k})=1 \label{constr1}
\ee

We associate to each hypersurface $M$ a weight $\exp(-H(M))/Z$,
where $Z=\sum_M \exp(-H(M))$ is the partition function of the system.
We suppose that the energy $H$ can be written as a sum of terms depending
on the configuration around each hyperedge, more precisely
\be
H^{d}_{gonihedric} = \sum_{\alpha_1<...<\alpha_{n+1},\vec r}
H_{\alpha_1...\alpha_{n+1}}(\vec r), \label{orham}
\ee
where $H_{\alpha_1...\alpha_{n+1}}(\vec r)$ depends on the $U$'s of the
hyperplaquettes by which the hyperedge is surrounded. These
are the $U$'s in eq. (\ref{constr1}).

In \cite{savvidywegner} the Hamiltonian counted the number of
right angles at all hyperedges (eq.(20) of \cite{savvidywegner}).
Here we will count the number of
hyperplaquettes $\Omega_{\alpha_1,...\alpha_n}(\vec r)$ in
$M_{d-n}$ which do not have a straight continuation at a hyperedge.
If we multiply this number by $2K$, then the contribution of a
hyperedge reads
\be
H_{\alpha_1...\alpha_{n+1}}(\vec r) = K \sum_k
(1-U_{\alpha_1...\alpha_{k-1}\alpha_{k+1}...\alpha_{n+1}}(\vec r)
U_{\alpha_1...\alpha_{k-1}\alpha_{k+1}...\alpha_{n+1}}
(\vec r-\vec e_{\alpha_k})).
\ee
In the following we will skip the unimportant constant. Then we may write
\be
H_{\alpha_1...\alpha_{n+1}}(\vec r) = -K \sum_k
V_{\alpha_1...\alpha_{k-1}\alpha_{k+1}...\alpha_{n+1},
\alpha_k}(\vec r).\label{ham1}
\ee
with\be
V_{\alpha_1...\alpha_{k-1}\alpha_{k+1}...\alpha_{n+1},
\alpha_k}(\vec r) =
U_{\alpha_1...\alpha_{k-1}\alpha_{k+1}...\alpha_{n+1}}(\vec r)
U_{\alpha_1...\alpha_{k-1}\alpha_{k+1}...\alpha_{n+1}}
(\vec r-\vec e_{\alpha_k}). \label{defV}
\ee
In this formulation the energy associated with a simple kink at a
hyperedge is $4K$ times the length of the kink.
As usual \cite{wegnermathphys} the constraint (\ref{constr1})
can be eliminated by the introduction of Ising spins $\sigma$ attached
to the $(d-n+1)$ dimensional hypercubes $\Omega_{\alpha_1...\alpha_{n-1}}$.
Then we obtain for $V$ the product over the spins at the boundary of two
neighbouring $(d-n)$-dimensional hyperplaquettes which are parallel
to each other
\begin{eqnarray*}
V_{\alpha_1...\alpha_{n},\beta}(\vec r) &=& \prod_{k=1}^n
\sigma_{\alpha_1...\alpha_{k-1}\alpha_{k+1}...\alpha_{n}}(\vec r)
\sigma_{\alpha_1...\alpha_{k-1}\alpha_{k+1}...\alpha_{n}}
(\vec r-\vec e_{\alpha_k}) \\
&&\sigma_{\alpha_1...\alpha_{k-1}\alpha_{k+1}...\alpha_{n}}
(\vec r-\vec e_{\beta})
\sigma_{\alpha_1...\alpha_{k-1}\alpha_{k+1}...\alpha_{n}}
(\vec r-\vec e_{\alpha_k}-\vec e_{\beta}).
\end{eqnarray*}

\subsection{Duality transformation}
First we describe the general algebraic procedure along the lines of
 \cite{wegnerphys} which allows to obtain the dual model of an Ising model.
For this purpose the Hamiltonian is written as a linear combination of
Ising variables as in (\ref{ham1})
\be
H=-\sum_i K_i V_i,
\ee
where for the moment we denote the products of the Ising variables by $V_i$.
They cannot assume values $\pm 1$ independently from each other,
for example due to the constraint (\ref{constr1}).
One has to find a complete set of constraints for the $V$'s so
that any set of spins $V$ which obeys these constraints yields at
least one allowed configuration of $U$'s.
These constraints have to have the form
\be
\prod_i V_i^{m_{ij}} = 1,~~~~m_{i,j}=0,1 . \label{constr}
\ee
Then the partition function can be written
\be
Z = 2^n \sum_{\{\sigma\},\{V\}} \prod_{i} V_i^{\sum_j m_{ij}\sigma_j}
e^{K_iV_i}.
\ee
With each constraint one variable $\sigma_j=0,1$ has been introduced.
If the constraint (\ref{constr}) is violated, that is the left hand-side of
this
equation equals $-1$ then the sum over $\sigma_j$ yields zero, but if this
fulfilled it yields twice the contribution. This factor of two is taken care of
in the prefactor $2^n$. Thus the summation over the $V$'s can be performed
independently without any constraints.

For a given configuration of the $V$'s there are in general $2^{n_d}$
configuration of the $U$'s.
Normally this is the degeneracy of the groundstate.
The exponent $n$ is the number $n_d$ minus the number of constraints.

With the $\sigma$'s we associate the Ising variables
\be
S_j=(-)^{\sigma_j}\label{dualS}
\ee
of the dual model.
Now the sum over each $V_i=\pm 1$ can be evaluated,
\be
e^{K_i} + e^{-K_i}\prod_j S_j^{m_{ij}} = \sinh(2K_i)
\exp(-K_i^{\star}\prod_j S_j^{m_{ij}})
\ee
with
\be
\exp(-2K_i^{\star})=\tanh(K_i). \label{Kdual}
\ee
Thus the partition function $Z$ is expressed
\be
Z=2^n \prod_i \sinh(2K_i) Z^{\star}
\ee
by the partition function $Z^{\star}$ of the dual model
\be
H^{\star}=-\sum_i K_i^{\star} \prod_j S_j^{m_{ij}}.
\ee

It may be, that not all of the constraints (\ref{constr}) are
chosen independently.
More precisely it may be, that they are multiplicatively dependent,
that is there exist numbers $n_j$, at least one of which is odd, so
that the sum $\sum_j n_j m_{ij}$ is an even number for all $i$.
Then the product of the conditions (\ref{constr}) with odd $j$ is
identically $1$, since each factor $V$ appears an even number of times.
In this case we obtain more Ising spins $S$ than necessary.
Then the dual Hamiltonian is invariant under the transformation
$S_j \rightarrow (-)^{n_j} S_j$.
If there are sets of $n_j$ which differ from zero only in a small region,
then this is a gauge transformation.

\subsection{Application to  hypersurfaces with linear action}

An obvious constraint is given by eq. (\ref{constr1}). It can
be written in terms of the $V$'s as
\be
\prod_{k=1}^{n+1} V_{\alpha_1...\alpha_{k-1}\alpha_{k+1}...\alpha_{n+1},
\alpha_k}(\vec r)=1. \label{constr2}
\ee
To find further constraints let us assume that we know the
$U(\vec r)$ for $\sum_{\alpha=1}^d r_{\alpha}<c$ which are
compatible with the $V$ and now we calculate the $U(\vec r)$
with $\sum_{\alpha=1}^d r_{\alpha}=c$ from those and the $V$'s.
Then apparently from the definition of $V$, eq. (\ref{defV}) we obtain
\be
U_{\alpha_1...\alpha_n}(\vec r)=
V_{\alpha_1...\alpha_n,\beta}(\vec r)
U_{\alpha_1...\alpha_n}(\vec r-\vec e_{\beta}).
\ee
For $n=d-1$ there is only one choice of $\beta$, since
the $\alpha$ and $\beta$ have all to be different.
For $n<d-1$ however we have different choices for $\beta$. Let
us choose $\gamma$ instead.
Then a solution requires
\be
V_{\alpha_1...\alpha_n,\beta}(\vec r) U_{\alpha_1...\alpha_n}
(\vec r-\vec e_{\beta})=
V_{\alpha_1...\alpha_n,\gamma}(\vec r)
U_{\alpha_1...\alpha_n}(\vec r-\vec e_{\gamma}).
\ee
This is equivalent to the condition
\be
V_{\alpha_1...\alpha_n,\beta}(\vec r)
V_{\alpha_1...\alpha_n,\gamma}(\vec r)
V_{\alpha_1...\alpha_n,\beta}(\vec r-\vec e_{\gamma})
V_{\alpha_1...\alpha_n,\gamma}(\vec r-\vec e_{\beta}) =1
\label{constr3}
\ee
We neglect any constraints due to the boundary condition.
We expect, that they do not contribute in the thermodynamic limit.

According to (\ref{dualS}) we introduce for each constraint an
Ising spin on the dual lattice.
We denote the Ising spin for the condition (\ref{constr2})
$\Gamma_{\alpha_1...\alpha_{n+1}}(\vec r)$.
Apparently it is invariant under the permutations of all $\alpha$'s, since
the condition (\ref{constr2}) has this property.
For the constraint (\ref{constr3}) we introduce the Ising spins
$\Lambda_{\alpha_1...\alpha_n,\beta\gamma}(\vec r)$, which is invariant under
the permutation of all $\alpha$'s and under the exchange of
$\beta$ and $\gamma$.
Now to each of the original terms
$K V_{\alpha_1...\alpha_{k-1}\alpha_{k+1}...\alpha_{n+1},
\alpha_k}(\vec r)$
there corresponds a term in the dual Hamiltonian
$H^{\star} \equiv K H^{d}_{dual}$.
To obtain this term we have to find all constraints in which
$V_{\alpha_1...\alpha_{k-1}\alpha_{k+1}...\alpha_{n+1},
\alpha_k}(\vec r)$
appears and to multiply the corresponding Ising spins on the dual lattice.
This yields the dual Hamiltonian
\be
H^{d}_{dual}
= - \sum_{\vec r}\sum_{\alpha_1<...<\alpha_n,\beta\neq\alpha_i}
\Gamma_{\alpha_1...\alpha_n\beta}(\vec r)
\prod_{\gamma\neq\alpha_i,\beta}
\Lambda_{\alpha_1...\alpha_n,\beta\gamma}(\vec r)
\Lambda_{\alpha_1...\alpha_n,\beta\gamma}(\vec r+\vec e_{\gamma}).
\label{hamd}
\ee

\subsection{Geometrical interpretation}

Thus we have to multiply $d-n-1$ factors $\Lambda(\vec r)$ with the
corresponding factors $\Lambda$ at sites $\vec r+\vec e_{\gamma}$ and
this interaction is mediated by an Ising spin
$\Gamma(\vec r)$. The spins
$\Gamma_{\alpha_1...\alpha_n\beta}(\vec r)$
can be considered to be located on the hyperedges
$\Omega_{\alpha_1...\alpha_n\beta}(\vec r)$.
They are bounded by $(d-n-2)$ dimensional hypervertex
$\Omega_{\alpha_1...\alpha_n\beta\gamma}(\vec r)$ and
$\Omega_{\alpha_1...\alpha_n\beta\gamma}(\vec r+\vec e_{\gamma})$ at which one
may consider the location of the Ising spins
$\Lambda_{\alpha_1...\alpha_n,\beta\gamma}(\vec r)$ which are multiplied by
$\Gamma$.
Note however, that at each of these hypervertex there is the location of
$(n+2)(n+1)/2$ different spins $\Lambda$.

Let us give a geometrical interpretation of the duality transformation
by comparing the high-temperature expansion of the dual model
(\ref{hamd}) with
the low-temperature expansion of the original model (\ref{orham}),
(\ref{ham1}),(\ref{defV}) .
In the high-temperature expansion one writes each factor
\be
\exp(K^{\star}\prod S) = \cosh(K^{\star})(1+\tanh(K^{\star})\prod S)
\ee
and expands in powers of $\tanh(K^{\star})$.
Only those
products over $S$ contribute to the sum over $S$ in which each $S$ appears an
even number of times.
Let us consider the products of $\Lambda_{\alpha_1...\alpha_n,\beta\gamma}$
from the dual
Hamiltonian (\ref{hamd})  for fixed $\alpha_1...\alpha_n$.
Each term consists of
$2(d-n-1)$ factors $\Lambda$ which are located at the boundaries of a
$(d-n-1)$-dimensional hypercube. In order
that each factor $\Lambda$ appears pairwise
each boundary of these hypercubes has to appear
for an even number of times.
Thus it constitutes a closed $(d-n-1)$-dimensional closed
hypersurface, which
encloses a flat $(d-n)$-dimensional hypersurface. At each of these
$(d-n-1)$-dimensional hypercubes there is a factor $\Gamma$.
They again have to
appear pairwise. Thus these $(d-n-1)$-dimensional hypercubes
are the edges of a
closed $(d-n)$-dimensional hypersurface. This hypersurface contributes to the
partition function with a term $\tanh(K^{\star})$ to the power twice the number
of elementary edge pieces. Thus the partition function agrees with that of
our original model provided
\be
(\tanh(K^{\star}))^2 = \exp(-4K)
\ee
in agreement with eq.(\ref{Kdual}).
For fixed $\Gamma$ the model
decays into models for spins $\Lambda$ with fixed $\alpha_1...\alpha_n$ in
$n$-dimensional hyperplanes with fixed coordinates $x_{\alpha_1}$ to
$x_{\alpha_n}$. The spins $\Gamma$ provide a coupling between these models.

\subsection{Gauge transformations}

In general the model will be gauge invariant.
Let us determine the number of gauge degrees of freedom per $d$-dimensional
elementary cell (hypercube).
The number $n_{iU}$ of independent variables $U$ per elementary cell is
\be
n_{iU} = {d-1 \choose n-1}
\ee
since for given $U(\vec r - \vec e)$ and $U_{1\alpha_2...\alpha_n}(\vec r)$
the other $U_{\alpha_1...\alpha_n}(\vec r)$ are obtained uniquely from the
constraint (\ref{constr1}). Thus $n_{iU}$ out of the
\be
n_V=(d-n){d \choose n}
\ee
variables $V$ are independent.
For the duality transformation $n_{\Gamma}$ and $n_{\Lambda}$ constraints are
used,
where $n_{\Gamma}$ and $n_{\Lambda}$ are the numbers of spins $\Gamma$ and
$\Lambda$ per
elementary cell,
\be
n_{\Gamma} = {d \choose n+1}, \mbox{    }
n_\Lambda=\frac{(d-n)(d-n-1)}2 {d \choose n}.
\ee
Thus
\be
n_g=n_{\Gamma}+n_{\Lambda}-n_V+n_{iU}
\ee
is the number of gauge degrees of freedom per elementary cell.

A first class of gauge transformations can be easily given for $n \le d-3$.
The simultaneous switching of the Ising spins
$\Lambda_{\alpha_1...\alpha_n,\beta\gamma}(\vec r)$,
$\Lambda_{\alpha_1...\alpha_n,\beta\delta}(\vec r)$,
$\Lambda_{\alpha_1...\alpha_n,\gamma\delta}(\vec r)$,
$\Lambda_{\alpha_1...\alpha_n,\beta\gamma}(\vec r-\vec e_{\delta})$,
$\Lambda_{\alpha_1...\alpha_n,\beta\delta}(\vec r-\vec e_{\gamma})$,
$\Lambda_{\alpha_1...\alpha_n,\gamma\delta}(\vec r-\vec e_{\beta})$
leaves the dual Hamiltonian invariant.
We denote this gauge transformation
by $G^{(1)}_{\alpha_1...\alpha_n,\beta\gamma\delta}(\vec r)$.
It is invariant under permutation of
the $\alpha$'s and separately of $\beta$, $\gamma$ and $\delta$.
These gauge transformations are not
independent from each other for $n \le d-4$, since the application of
the following eight
transformations leave all spins invariant, \begin{eqnarray*}
G^{(1)}_{\alpha_1...\alpha_n,\beta\gamma\delta}(\vec r)
G^{(1)}_{\alpha_1...\alpha_n,\beta\gamma\epsilon}(\vec r)
G^{(1)}_{\alpha_1...\alpha_n,\beta\delta\epsilon}(\vec r)
G^{(1)}_{\alpha_1...\alpha_n,\gamma\delta\epsilon}(\vec r) &&\\
G^{(1)}_{\alpha_1...\alpha_n,\beta\gamma\delta}(\vec r-\vec e_{\epsilon})
G^{(1)}_{\alpha_1...\alpha_n,\beta\gamma\epsilon}(\vec r-\vec e_{\delta})
G^{(1)}_{\alpha_1...\alpha_n,\beta\delta\epsilon}(\vec r-\vec e_{\gamma})
G^{(1)}_{\alpha_1...\alpha_n,\gamma\delta\epsilon}(\vec r-\vec e_{\beta})
&=&1.  \label{inv1}
\end{eqnarray*}
These gauge transformations are independent, if one chooses for given
$\alpha$'s
the smallest possible $\beta$.
Those with larger $\beta$'s can be constructed from the first ones by
use of eq. (\ref{inv1}). Thus one has
\be
n_{g1}=\frac{(d-n-1)(d-n-2)}2 {d \choose n}
\ee
independent gauge degrees of freedom from $G^{(1)}$.

A second class of gauge transformations denoted by
$G^{(2)}_{\alpha_1...\alpha_{n+2}}(\vec r)$ exists for $n \le d-2$
and involves also the $\Gamma$'s.
It is invariant under all permutations of the
$\alpha$'s.
This transformation switches the spins
$\Lambda_{\alpha_1...\alpha_{k-1}\alpha_{k+1}...\alpha_{l-1}\alpha_{l+1}
...\alpha_{n+2},\alpha_k\alpha_l}(\vec r)$ for all $k<l$,
$\Gamma_{\alpha_1...\alpha_{k-1}\alpha_{k+1}...\alpha_{n+1}}
(\vec r - \vec e_{\alpha_k})$ and
$\Gamma_{\alpha_1...\alpha_{k-1}\alpha_{k+1}...\alpha_{n+1}}
(\vec r )$ for all $k$.
 Again these gauge transformations are not independent
for $n \le d-3$, since
\begin{eqnarray*}
\prod_{k=1}^{n+3}
G^{(2)}_{\alpha_1...\alpha_{k-1}\alpha_{k+1}...\alpha_{n+3}}(\vec r)
G^{(2)}_{\alpha_1...\alpha_{k-1}\alpha_{k+1}...\alpha_{n+3}}
(\vec r-\vec e_{\alpha_k}) &&\\
\prod_{k<l<m}
G^{(1)}_{\alpha_1...\alpha_{k-1}\alpha_{k+1}...\alpha_{l-1}\alpha_{l+1}...
\alpha_{m-1}\alpha_{m+1}...\alpha_{n+3},\alpha_k\alpha_l\alpha_m}(\vec
r) &=& 1. \label{inv2}
\end{eqnarray*}
By means of this equation we can express all
$G^{(2)}_{\alpha_1...\alpha_{n+2}}(\vec r)$ with $\alpha_i >1$ for all $i$
 by those where one $\alpha$ equals $1$ and
$G^{(2)}(\vec r-\vec e)$ and by $G^{(1)}$'s.
The number of $G^{(2)}$'s with $\alpha_1=1$ per lattice site is
\be n_{g2}={d-1 \choose n+1}.
\ee
They are independent gauge transformations. Since
\be
n_g = n_{g1}+n_{g2}.
\ee
these two types of gauge transformations exhaust all independent
gauge transformations.

Now we will consider some special cases.

\subsection{Two-dimensional surfaces}

In the case $d-n=2$ we have two-dimensional surfaces $M_{2}$.
If we denote the $\Gamma_{\alpha_1...\alpha_{d-2}\beta}(\vec r)$
by $\Gamma(\vec r,\vec r +\vec e_{\gamma})$ where
$\gamma$ is the index, that is different from the $\alpha_i$
and $\beta$, and the $\Lambda_{\alpha_1...\alpha_{d-2},\beta\gamma}$
by $\Lambda_{\beta\gamma}$, then the Hamiltonian reads
\be
H^{d}_{dual} = -\sum_{\vec r}\sum_{\beta \neq \gamma}
\Lambda_{\beta\gamma}(\vec r)
\Gamma(\vec r,\vec r +\vec e_{\gamma})
\Lambda_{\beta\gamma}(\vec r+\vec e_\gamma).
\ee
For fixed $\Gamma$ this model can be regarded as a model
of two-dimensional Ising-models in plains parallel to
the $\beta,\gamma$-plain. The $\Gamma$'s produce a coupling between
these Ising-models.
Here we have one gauge degree of freedom
$G^{(2)}_{1...d}(\vec r)$ per elementary cell.
Indeed the dual Hamiltonian is invariant under simultaneous switching
of all $\Lambda_{\beta\gamma}(\vec r)$,
$\Gamma(\vec r - \vec e_{\gamma},\vec r)$
and $\Gamma(\vec r , \vec r + \vec e_{\gamma})$
for fixed $\vec r$.

\subsection{Hypersurfaces of codimension one}

In this case of a $d-1$ dimensional hypersurface $M_{d-1}$
the constraint (\ref{constr2}) reads
\be
V_{\alpha,\beta}(\vec r)V_{\beta,\alpha}(\vec r)=1,
\ee
thus $V_{\alpha,\beta}(\vec r)=V_{\beta,\alpha}(\vec r)$.
If one considers these two $V$'s to be identical,
then the constraint (\ref{constr2}) and the Ising spin $\Gamma_{\alpha\beta}$
disappears. Then the original Hamiltonian reads
\be
H = -2K \sum_{\alpha<\beta,r} V_{\alpha,\beta}(\vec r)
\ee
and $V_{\alpha,\beta}(\vec r)$ does not only appear in the constraint
associated with $\Lambda_{\alpha,\beta\gamma}(\vec r)$ and
$\Lambda_{\alpha,\beta\gamma}(\vec r+\vec e_{\gamma})$ but also
in the conditions for $\Lambda_{\beta,\alpha\gamma}(\vec r)$
and $\Lambda_{\beta,\alpha\gamma}(\vec r+\vec e_{\gamma})$.
This yields
\be
H_2^{\star} = -K_2^{\star} \sum_{\alpha<\beta,r}\prod_{\gamma}
\Lambda_{\alpha,\beta\gamma}(\vec r)
\Lambda_{\alpha,\beta\gamma}(\vec r+\vec e_{\gamma})
\Lambda_{\beta,\alpha\gamma}(\vec r)
\Lambda_{\beta,\alpha\gamma}(\vec r+\vec e_{\gamma})
\ee
with
\be
\exp(-2K_2^{\star})=\tanh(2K).
\ee
This expression can also be obtained from $H^{\star}$ by
summing over the $\Gamma$
\be
\exp(-H_2^{\star})=\mbox{const.}\sum_{\Gamma} \exp(-H^{\star})
\ee
where use is made, that a given $\Gamma$ appears only in two
terms $\Gamma \Lambda^{2(d-n-1)}$ in $H^{\star}$.
This elimination of $\Gamma$ is a modification of the decoration
transformation \cite{deco}.

\subsection{Hypersurfaces of codimension two}

In this case of a $d-2$ dimensional hypersurface $M_{d-2}$ a given
$\Gamma_{\alpha_1\alpha_2\alpha_3}(\vec r)$
is multiplied by the three terms
\be
A_i=\prod_{\gamma}(\Lambda_{\alpha_i\alpha_{i+1},\alpha_{i+2}\gamma}(\vec r)
\Lambda_{\alpha_i\alpha_{i+1},\alpha_{i+2}\gamma}(\vec r+\vec e_{\gamma})).
\ee
Here the indices $i+1$ and $i+2$ are meant modulo 3.
Then summation over the $\Gamma$'s which is analogous to the
star-triangle transformation \cite{startri} yields
\be
-H_3^{\star} = K_3^{\star} \sum (A_1 A_2 + A_1 A_3 + A_2 A_3).
\ee
with
\be
\exp(2K_3^{\star}) = \exp(2K^{\star}) -1 + \exp(-2K^{\star}).
\ee
For larger values of $n$ one can also eliminate $\Gamma$. Then however the new
interaction involves not only bilinear terms in the $A$'s, but
also terms of higher order in the $A$'s.

\subsection{Random walks}

In the case $d-n=1$ we have the model of closed random walks $M_{1}$.
There are no variables $\Lambda$, since there is no constraint (\ref{constr3}).
Then there is only one Ising spin $\Gamma$ at each lattice site
appearing in $d$ terms.
The dual Hamiltonian simply reads
\be
H^{\star} = -K^{\star}d \sum_{\vec r} \Gamma(\vec r).
\ee
Thus there is no interaction between the spins.
They are only subject to an external field $K^{\star}d$.
Obviously there is no phase transition in this case.
The partion function reads
\be
Z^{\star} = (2 \cosh(dK^{\star}))^N
\ee
where $N$ is the number of elementary cells.
In this model there is no gauge invariance.

\vspace{1cm}
{\Large{\bf Conclusion and Acknowledgements}}
\vspace{.5cm}

In this paper we describe the lattice implementation of the string with
linear action which allows self-intersections.
For this model not only an equivalent Ising-spin model is given,
but also a dual model was constructed.
The generalization for $p$-membranes is given.
In general the models have gauge degrees of freedom.

We should mention that the dual representations are
important in our attempt to understand confinement
phase of the gauge theory \cite{thooft,mandelstam,fradkin,seiberg}.
Recent results in this direction can be found in
\cite{luis,veneziano}.

One of the authors (F.J.W.) gratefully acknowledges, that part of this
work was supported by the Foundation for Research and Technology-Hellas
and (G.K.S) is supported in part by the Alexander von Humboldt
Foundation.

\vspace{1cm}
{\it Note added}
\vspace{.5cm}

We are thankful to the referee for pointing out an important
article of A.Cappi, P.Colangelo, G.Gonnella and A.Maritan
in Nucl.Phys. B370 (1992) 659 ,where the authors consider
the general three-dimensional
spin model with straight, diagonal and square spin interaction.
In terms of interface surfaces the model contains an area term
$\beta_s$, the gonihedric term $\beta_c$ and self-intersection
term $\beta_l$.

The special case, which we discuss in three
dimensions, corresponds to $\beta_s = 0$, $\beta_c =1$ and
$\beta_l = 4k-2$, with $k=0$.
The principle point is that $\beta_s = 0$ and the case $k=0$
corresponds to the
"super symmetric" point (0,1,-2) in the space of Hamiltonians
$(\beta_s ,\beta_c , \beta_l)$, because at that point the
degeneracy of the vacuum is extremely high and is equal to
$2^{dN}$ for the lattice of the size $N^d$.
The curvature representation shows that
the systems with and without area term are in different class of
universality (see \cite{savvidy4} and  section $1.4$).
String interpretation of this spin systems requires the existence
of the second order phase transition and an appropriate continuum limit
at the critical point.

Second remark concerns the non-Abelian extension of the two-plaquette
action (\ref{4Dham}). If we exchange the spin variable $\sigma$ by
the element of the $SU(N)$, $U=exp(iga A)$, then the Hamiltonian
takes the form
\be
H= -\sum_{pairs~of~parallel~plaquetts}\frac{\beta}{n^2}
(Tr U_{\Box} + Tr U^{\star}_{\Box})(Tr U_{\Box} + Tr U^{\star}_{\Box})
\ee
and in the naive continuum limit reproduces the usual Yang-Mills action.

\end{document}